\documentclass[%
9pt,
 aip,
rsi,%
 amsmath,amssymb,
preprint,%
]{revtex4-1}

\usepackage{graphicx}
\usepackage{dcolumn}
\usepackage{bm}
 \graphicspath{ {plots/} }
 
 \usepackage[export]{adjustbox}
 \usepackage{ulem}
 \usepackage{anyfontsize}
 \usepackage[textwidth=1.2cm,textsize=tiny]{todonotes}
\newcommand{\thermalconductivity}{$\mathrm{W m^{-1} K^{-1}}$}
\newcommand{\hcoefficient}{$\mathrm{W m^{-2} K^{-1}}$}
\newcommand{\mobility}{$\mathrm{m^{2} V^{-1} s^{-1}}$}

\begin{document}

\preprint{aip}

\title[Heat transport in OLEDs]{Impact of thermal transport parameters on the operating temperature of OLEDs}

\author{G. Krikun}
\affiliation{ 
Institute of Solid State Physics and NAWI Graz, Graz University of Technology, Graz, 8010, Austria
}%
\author{K. Zojer}%
 \email{karin.zojer@tugraz.at}
\affiliation{ 
Institute of Solid State Physics and NAWI Graz, Graz University of Technology, Graz, 8010, Austria
}%

\date{\today}

\begin{abstract}
Excess heat in organic light emitting diodes (OLEDs) that is produced during their operation may accelerate their degradation and may cause an inhomogeneous brightness distribution, in particular in large area OLEDs.
Assessing the quantitative impact of heat excess is difficult, because all decisive processes related to charge transport and emission via charge recombination are thermally activated.
For example, electric currents that are elevated due to larger temperatures cause additional Joule heating and increase, hence, the device temperature even further. Here we establish how parameters responsible for heat transport, i.e., the thermal conductivity of the organic layers and the heat transfer coefficient between the device surface and the environment govern the temperature inside the OLED. 
We establish, relying on three-dimensional drift-diffusion simulations that self-consistently couple thermally-activated charge transport and heat transport, that the  thermal conductivity of organic layers is not a bottleneck for heat transport, because the encountered layer thicknesses in realistic device geometries prevent heat accumulation.
The heat transfer to the ambient environment is the key parameter to dissipate excess heat from the device. 
Intentionally elevated operating temperatures, that may improve the OLEDs electric performance, are not necessarily beneficial, as any increase in operating temperature decreases the device stability. 
The thermal effects being decisive for the OLED temperature occur in device layers beyond the electrically active region.
We propose analytical expressions that relate the temperature in the device for a given point of operation to the heat transfer to the environment and the substrate.
\end{abstract}

\maketitle

\section{Introduction}

Two major factors presently hinder the application of organic light emitting diodes (OLEDs) for large-area lighting applications: device degradation and inhomogeneous light emission from the exterior surface.\cite{Park2011}
Both factors strongly relate to excess heat that is generated during operation.
On the one hand, elevated temperatures inside the OLEDs accelerate degradation,\cite{Xu2004,Zhou2000,Nenna2007} alongside with other factors such as water penetrating into the device \cite{Li2016}, surface roughness, and mass diffusion of material.
On the other hand, local variations in temperature and, hence, in the in-plane current density give rise to an inhomogeneous brightness distribution.\cite{Garditz2007,Kohari2013} 
The primary origin of such locally elevated current densities and temperatures are non-uniformly distributed driving voltages, because these voltages differ due to position-dependent ohmic losses in the large area contacts.\cite{Fischer2014} 
As a consequence, the efficiency of how heat is dissipated from the device becomes dependent on the position in which the heat is generated. 
Even if heating is compensated by heat outflow,\cite{Park2014} such instabilities may lead to local "bright spots" in which the device will degrade much faster.

The quantitative impact of heat excess is particularly difficult to assess, because the major effects that govern the electrical performance are thermally activated, i.e., the charge transport and the recombination efficiency in the organic layers. 
An increasing temperature enlarges the charge carrier mobility in organic thin films\cite{Bassler1993} and the recombination rates and, hence, increases the device current. In turn, this enhanced current causes larger Joule heating and elevates the temperature even further.
If this process of self-activated heating is balanced by a sufficient outflow of heat, the current reaches a stable steady state. 
Unbalanced, the self-activated heating process can drive the device into a ''vicious cycle'', which will eventually overheat the device to the point where it is not operable anymore.

In this article, we pursue the question how the parameters that are responsible for heat transport in OLEDs govern the temperature distribution in the device.
Inspired by the vast efforts spent to optimize the material properties for charge transport, our article focuses on the \textit{thermal properties} that most favorably support the device operation. 
Specifically, we seek to learn (i) whether the parameters that govern heat transport represent separate tuning handles to dissipate heat, and (ii) with which layers of the OLED stack the heat dissipation can be efficiently controlled.
As further asset we seek to determine a relation between the device temperature, the electric current, and the heat transport parameters.
With such a relation, even if approximate, we would be able to predict how the OLED operation is affected by its thermal transport properties and able to estimate the impact of heat dissipation directly from an experimentally obtained current voltage characteristics.

\begin{figure}
    \centering	
    \includegraphics{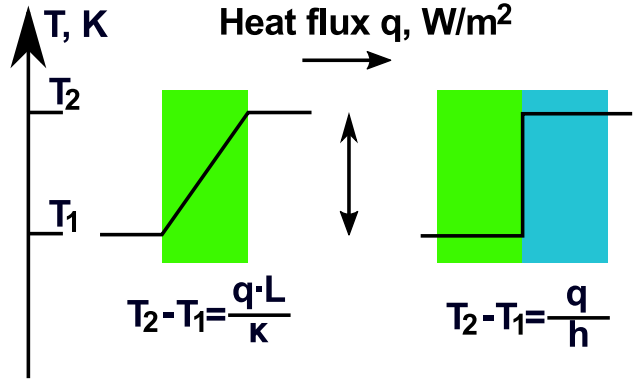}
    \caption{ Schematic illustration of the thermal conductivity $\kappa$ (left) and the thermal transport coefficient $h$ (right). Without heating in the layer, the thermal conductivity $\kappa$ is the coefficient of proportionality between a heat flux $q$ and the temperature gradient $T_2 - T_1$ (Fourier law), $L$ denotes the thickness of the layer. For a heat flux $q$ across an interface formed by two materials, the heat transfer coefficient $h$ determines the difference between the temperatures on each side of device.}
    \label{fig:bcpic}
\end{figure}

As a first parameter, we consider the thermal conductivity $\kappa$ of individual organic layers. 
As illustrated in Figure~\ref{fig:bcpic}, $\kappa$ describes how efficient heat is transported from the heat source through a layer of a thickness $L$. 
The second considered parameter is the heat transfer coefficient $h$ that describes heat transport through an interface (Figure~\ref{fig:bcpic}).\cite{Jin2014}
In the context of OLEDs, the heat transfer coefficient between the outermost layers and the ambient environment  appears to be most crucial.
There are clear indications that the heat outflow from the device can be sufficiently improved by choosing an appropriate substrate.\cite{Chung2009,Triambulo2016} 
As the generated heat has to be dissipated with respect to ambient temperature, the absolute temperature within the organic layers will be influenced not only by the electric current, but also by the ambient temperature.

We employ drift-diffusion based simulations to monitor how these two types of parameters, i.e., $\kappa$ and $h$, contribute to the thermal and electrical behavior of the OLED.
With such simulations, we simultaneously account for charge and heat transport based on the (i) geometry of the OLED stack, (ii) the material parameters governing the charge and the thermal transport, and (iii) the coupling between charge and heat transport to establish a self-consistent feedback between them.
This allows us to obtain current-voltage characteristics and temperature-voltage relation for an OLED that is operated at a given ambient temperature.

\section{Methodology}

\begin{figure}
	\centering
    \includegraphics{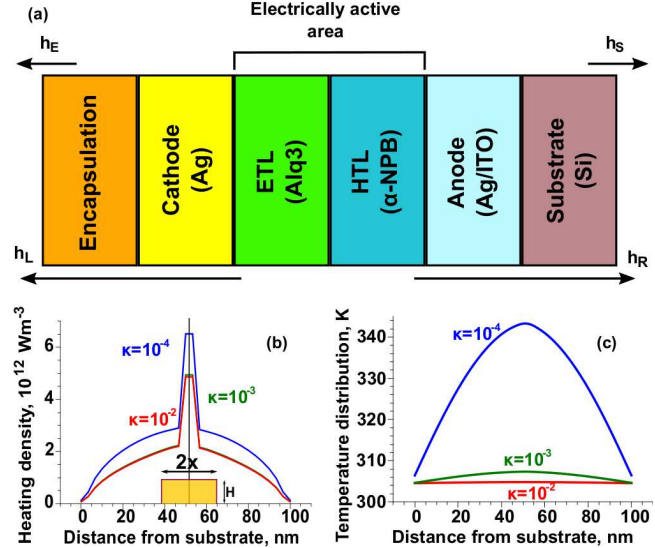}
    \caption{(a) Schematic cross section of an OLED containing an electron and hole conducting layer sandwiched between two contacts. 
    Beyond these electrical contacts, an encapsulation layer and a substrate with their heat transfer coefficients to the ambient, $h_{\mathrm{E}}$ and $h_{\mathrm{S}}$, are considered. 
    Within the organic layers, the coupled charge and heat transport is self-consistently simulated with a drift-diffusion approach. 
    The combined impact of heat transfer to the ambient, heat transfer between and through subsequent encapsulation or substrate layers can be lumped into an effective heat transfer coefficient %
    $h_{\mathrm{L}}$ (encapsulation side) and $h_{\mathrm{R}}$ (substrate side), respectively.
    (b) Profile of the heat density distribution in a symmetric device between the electrical contacts that are separated by $L$~=~50~nm, no injection barrier,  $h$~=~$h_{\mathrm{L}}$~=~$h_{\mathrm{R}}$~=~$10^{4.25}$~\hcoefficient, $\kappa=10^{-2}$(bottom)/$10^{-3}$(middle)/$10^{-4}$(top)~\thermalconductivity~ in each organic layer operated at room temperature $T$~=~300~K with $V$~=~5~V. 
    (c) Corresponding temperature distribution.}
\label{fig:setup}
\end{figure}

To efficiently model the charge and heat transport, we distinguish between two device regions, i.e., (i) the region in which heat is generated, and (ii) in device regions, in which solely heat is transported. 
These regions are reflected in the schematic OLED cross section in Figure~\ref{fig:setup} as follows:
In the organic layers between the electric contacts, the electric current produces both Joule heating and heating due to non-radiative recombination. 
Hence, the region comprising the organic layers and the contacts will be described with a three dimensional drift-diffusion model for coupled charge and heat transport.

In the regions outside the contacts, i.e., in the encapsulation and substrate layers, no further heat is produced. 
In the absence of electric current, it is sufficient to solve the heat transport equation.
The latter equation can be solved analytically for steady state operation, sparing us to solve the heat transport equation for the entire device with the  elaborate numerical approach necessary for the electrically active region.
For convenience, the OLED cross-section shown in Figure~\ref{fig:setup} is oriented such that the layers are stacked along the z-axis and laterally extended in the x-y plane.
In this orientation, the externally applied voltage shall be uniformly applied between cathode and anode, i.e., the voltage drops in z-direction and is uniform in x and y directions.

\subsection{Drift diffusion model} 
The drift-diffusion model is an established approach for the investigation of the electrical performance of OLEDs.\cite{Davids1998,Knapp2010,Knapp2011,Ruhstaller2003,Slawinski2011,VanMensfoort2008} 
Since the model describes transport processes with effective material parameters, the model can be straight-forwardly extended to consider also heat transport and thermoelectric effects.\cite{Park2011a}

All charge densities, energy densities, and material parameters are dependent on the position in the device, %
so that the fluxes of the thermal energy and charges are three component vectors.
Below we describe the formulation and the implementation of the fully three dimensional drift-diffusion model.
Such a consideration of three-dimensional fluxes and related density distributions is crucially required to address lateral variations in temperature and in-plane current densities. 
Hence, we are aiming at devising a simulation code with which one can address temperature-related inhomogeneity issues that are beyond the scope of this work. 

In general, charge and heat transport are governed by two types of equations, the continuity equations and the current density equations.
The continuity equations for the densities of mobile charges, i.e., of electrons, $n$, and holes, $p$, and the density of thermal energy, $\rho c_p T$, read:

\begin{subequations}
	\begin{gather}
		\frac{\partial p}{\partial t} - \mathrm{div}\left( \mathbf{j_p}\right) =G-R\label{eq:contp}\\	
		\frac{\partial n}{\partial t} - \mathrm{div}\left( \mathbf{j_n}\right)=G-R\label{eq:contn}\\
		\rho c_p \frac{\partial T}{\partial t}-\mathrm{div}\left(\mathbf{q_T}\right)=H\label{eq:contT}\qquad 
	\end{gather}
    \label{eq:cont}
\end{subequations}

Therein, $\mathbf{j_n}$ and $\mathbf{j_p}$ are the electron and hole flux (rather than the electric current densities) respectively, $G$ and $R$ the generation and recombination rates. We are using the Langevin recombination model.\cite{Langevin1903} 
Note that there are more refined models to account for recombination rates which, however, would introduce additional free parameters to the our description.\cite{Lee2014,VanDerHolst2009} 
In Eq.~\ref{eq:contT}, $T$ refers to the temperature, $\rho$ to the mass density, $c_p$ to the specific heat capacity, and  $\mathbf{q_T}$ to the heat flux.
The term $H$ accounts for the amount of heat energy added or abstracted per unit volume and per unit time. 
The current density equations connect the fluxes with the densities: 
\begin{subequations}
	\begin{gather}
		\mathbf{j_p}=\mu_p p \left(\mathbf{\nabla} \phi + S \mathbf{\nabla} T \right) - D_p \mathbf{\nabla} p \label{eq:ddjp} \\
		\mathbf{j_n}=-\mu_n n \left(\mathbf{\nabla} \phi + S \mathbf{\nabla} T \right) - D_n \mathbf{\nabla} n \label{eq:ddjn} \\
		\mathbf{q_T}=-\kappa \mathbf{\nabla} T\qquad \label{eq:fourierlaw}
	\end{gather}
    \label{eq:connectionDensitiesToFluxes}
\end{subequations}
Eqs.~(\ref{eq:ddjp},\ref{eq:ddjn}) contain the material parameters responsible for charge transport, i.e., $\mu_i$ is the mobility of the corresponding charge carriers, $D_i$ the diffusion constant, and $S$ the Seebeck coefficient. Eq.~(\ref{eq:fourierlaw}) formulates the Fourier law and contains the heat conductivity $\kappa$.
To account for the hopping nature of the transport in organic materials, we consider mobilities and a Seebeck coefficient that  depend on electric field, temperature, and charge carrier concentration. 
There are different methods to incorporate this dependencies,\cite{Fishchuk2010,Fishchuk2013} we use relations that were previously extracted from Monte Carlo simulations.\cite{Lu2015,Pasveer2005}

The current densities $\mathbf{j_p}$ and $\mathbf{j_n}$ across organic-organic heterojunctions are also treated with Eqs.~(\ref{eq:ddjp},\ref{eq:ddjn}).
Besides drift and diffusion, we introduce an additional driving force for electrons and for holes at the position of the heterojunction. 
The extent of this force reflects how the transport-relevant levels of the two organic materials align at the heterojunction.
We obtain this force from a generalized electrostatic potential, $\tilde{\phi}$, that contains two additional terms compared to the actual electrostatic potential $\phi$. 
The first term accounts for the difference in the nominal transport level energies $E_{\mathrm{HOMO}}$ and $E_{\mathrm{LUMO}}$ of the two materials by introducing an additional potential drop across the heterojunction, i.e., $E_{\mathrm{HOMO}_1}-E_{\mathrm{HOMO}_2}$ for holes and $E_{\mathrm{LUMO}_1}-E_{\mathrm{LUMO}_2}$ for electrons, respectively.\cite{Sutherland}
The second term accounts for the fact that the density of states available for hopping transport is changing across the heterojunction.
We cast this change in the density of states into a local contribution to the generalized electrostatic potential following the argumentation in Ref.~\cite{Liemant}.
The current densities $\mathbf{j_p}$ and $\mathbf{j_n}$ entering and leaving the OLED across the contacts are described by a net injection current.
This injected current is a superposition of different contributions that are associated to the offset between the transport levels and the workfunction of the electrode metal, forming the injection barrier. We specifically consider current density contributions due to thermionic and tunneling injection (from contact into organic layer) as well as interface recombination (from organic layer into the contact).\cite{Davids1998a,Gruber2012,Scott1999}
The individual contributions are self-consistently determined as a function of the local situation present at each contact during operation. Particularly important are the temperature (thermionic injection) and the local electric field (thermionic and tunneling injection). 

The local electric field that drives the charges is given as the gradient $-\mathbf{\nabla}\phi$ of the electrostatic potential $\phi$. 
At heterojunctions, this field is replaced by an effective field $-\mathbf{\nabla}\tilde{\phi}$ related to the generalized electrostatic potential $\tilde{\phi}$. 
The potential $\phi$ is connected to the local net charge density via the Poisson equation 
\begin{equation}
    \epsilon_0 \mathbf{\nabla} \left( \epsilon_r \mathbf{\nabla} \phi \right) = q \left(n-p\right), \label{eq:poisson}
\end{equation}
in which $\epsilon_r$ is relative dielectric permittivity and $\epsilon_0$ the vacuum permittivity.
The potential set at the cathode corresponds to the applied voltage $V$, while the potential at the anode is set to zero. 

The heat flow $\mathbf{q_T}$ across the organic-organic heterojunction is considered in Eq.~(\ref{eq:fourierlaw}) via a thermal conductivity $\kappa$ that is averaged with respect to the $\kappa$ of the layers meeting at the interface.
The heat flow is directed out of the device, i.e., $\mathbf{q_T} = \mathbf{q}_z > 0$ at $z=L$,  if the temperature in the device exceeds the ambient temperature.
The heat flow out of the contacts, i.e., the flow leaving the region in which heat is generated, is assumed to be proportional to the difference in the temperature of the contact surface, e.g., at $z=L$, and the ambient environment $T_{\mathrm{ambient}}$:
\begin{equation}
	\label{eq:bcheatflux}
	\mathbf{q}_z\left(z=L\right)=h_{R}\left(T_{\mathrm{ambient}}-T\left(z=L\right)\right)
\end{equation}
Therein, an effective heat transfer coefficient $h_R$ dictates, how well the contact at $z=L$ is thermally connected to the ambient environment. 
Analogously, we also assume an effective heat transfer coefficient $h_L$ for the contact surface at $z=0$ (Fig.~\ref{fig:bcpic}).
The higher the heat flux through the contact interface is, the higher is the temperature gap $T_{\mathrm{ambient}}-T_{\mathrm{contact}}$ between the contact surface and the ambient.
The effective heat transfer coefficients $h_L$, $h_R$ are directly related to the propagation of heat through the subsequent layers between the respective contact and the environment. 
As will be explained in the section "Analytic solution of the heat equation" below, we can express $h_{L}$ and $h_{R}$ in terms of all thermal layer properties that connect the contact surfaces
to the ambient environment, i.e., the $\kappa$ of the individual layers, the coefficients $h$ between these layers, and the heat transfer coefficients $h_{E}$ and $h_{S}$ between the outermost encapsulation and substrate layer to the ambient.
With this definition, $h_L$, $h_R$ enable us to monitor directly the heat exchange between the electrically active layers and the ambient environment acting as heat sink.

The system of equations (\ref{eq:cont},\ref{eq:poisson}) with the above-stated boundary conditions is solved in the regions of the organic semiconducting layers. 
Periodic boundary conditions are imposed along the spatial directions perpendicular to the stacking direction.
We discretized the equations (\ref{eq:cont},\ref{eq:poisson}) with the Finite Differences Method on a non-uniform rectangular 3D mesh (for drift-diffusion equation we use the Scharfetter-Gummel discretization scheme \cite{SG}) and solved them iteratively with an explicit time discretization scheme (Euler forward) until steady state was achieved.
The solving algorithm was implemented in Fortran 95 and uses Open MPI to parallelize the computation. 

For our problem at hand, it is sufficient to consider only one spatial dimension, i.e., using a 1x1xN grid to discretize the cross-section shown in Figure~\ref{fig:setup}-a. 
This is, because here the potential drop across the device is assumed to be independent from the lateral position in the contact surface.
We carefully tested multiple devices with varying lateral extensions to ensure that the results from the one-dimensional simulations coincide with the ones obtained on a three-dimensional grid.
Furthermore, we verified that the simulations converged to a steady state that did not depend on the initial conditions assumed in the simulations.

\subsection{Analytical solution to the heat transfer equation}
\begin{figure}
	\centering
	\includegraphics[width=0.4\textwidth]{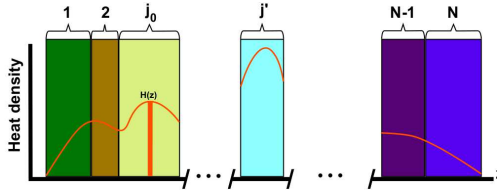}
	\caption{Discretized stack of multiple layers for solving the 1D heat equation. The line indicates the heat density distribution $H(z)$. Top labels enumerate individual slices in the stack, each of which has its own thermal conductivity $\kappa_{j'}$ and thickness $L_{j'}$. }
	\label{fig:MultiLayersSetup}
\end{figure}

Outside of the electrical contacts, the temperature distribution $T(z)$ corresponds to the solution of the steady-state heat equation. 
Since the voltage is constant in both lateral directions, i.e., perpendicular to the stack axis, we anticipate a temperature variation only along the stack. The temperature distribution for steady state operation is correspondingly given by the one-dimensional, the steady-state heat equation (cf. Eq.~(\ref{eq:contT})):
\begin{equation}
	\label{eq:1Dheatequation}
    \frac{d}{dz}\left(\kappa(z) \frac{dT(z)}{dz}\right) = -H(z).
\end{equation}
The source term on the right hand side contains the heat generation profile $H(z)$.
The general analytic solution $T(z)$ of Eq.~(\ref{eq:1Dheatequation}) can be obtained with a Greens function approach for an arbitrary layer sequence and for an arbitrary, static heat generation profile. 
Such a sequence is indicated in Figure~\ref{fig:MultiLayersSetup}.
Note that Eq.~(\ref{eq:1Dheatequation}) can be solved in terms of discretized slices being considerably thinner than the actual layers. However, such a refined discretization is usually necessary only if the heat generation profile is non-uniform.
The full expressions for the solution $T(z)$ are given in the Supporting Information. 

With the general steady-state solution at hand, we can construct the effective heat transfer coefficients $h_{L}$ and $h_{R}$ that represent the thermal transport behavior of all layers between the ambient environment and the electrical contacts. $h_{L}$ and $h_{R}$ will serve as boundary conditions for the heat equation that is solved in the region between the contacts.
Since we presume that the heat is exclusively generated in the region between the contacts and that this heat flows out of the device, we can interpret $h_{L/R}^{-1}$ as the total thermal resistance due to all layers the thermal flux had to pass through on its way to respective edge of the device.
Such substrate- or encapsulation-sided thermal resistances consist of the sum of all individual thermal resistances that are located between the contact and the ambient environment, be them due to (i) the heat transfer coefficient $h_{i,j}^{-1}$ at each interface between layers $i$ and $j$ or due to (ii) the thermal conductivities ${L_i}/{\kappa_i}$ in each layer $i$.  
Recalling that $h_{E/S}^{ext}$ is the thermal transport coefficient of the interface between the outermost encapsulation (/substrate) surface and the ambient environment, we get:
\begin{equation}
\frac{1}{h_{L/R}} = \frac{1}{h_{E/S}^{ext}} + \sum \frac{L_i}{\kappa_i}.
\label{eq:effectiveh}
\end{equation}

\subsection{Device layout}
A major challenge for the simulation of the transport through the layers and across the layer interfaces is the large number of required parameters.
Each layer requires the heat conductivity $\kappa$ and a set of parameters that describe charge transport. 
Each pair of interfaces requires parameterized conditions that ensure charge and energy conservation, such as heat transfer coefficients and injection currents. 
An unbiased systematic variation of all these parameters yields only limited insight and does not provide leads for relevant relations without generating a very large set of data.
Therefore, we seek a model device for our simulations, that serves three purposes: 
(i) The device layout should support the setup of a toy model with which the role of the heat transfer coefficients $h$ towards the environment and the heat conductivities $\kappa$ of the organic layers can be conveniently distinguished.
(ii) The toy model should operate with a strongly reduced amount of parameters so that the impact of charge and heat transport can be disentangled best.
(iii) It must be possible to straight-forwardly extent our toy model to accommodate more involved device structures with our modeling methodology.

Figure~\ref{fig:setup}-a shows a cross-section of a suitable device. 
While state-of-the-art devices contain a large number of distinct layers, we explicitly consider here two organic layers in which either predominantly hole or electron transport takes place. 
The two organic semiconducting layers are sandwiched between an anode and a cathode.
The organic layer adjacent to the cathode assumes the role of a electron transport layer while the one adjacent to the anode preferentially transports holes.
The division into two organic layers that form a heterojunction allows us to account for multiple conceivable charge transport scenarios in an effective manner.
Elaborate stacks for encapsulation or substrates are accounted for in this cross-section by one substrate and one encapsulation layer to capture their characteristic thickness and thermal properties. In general, however, the thermal behavior of the full sub-stacks that are not electrically active can be cast into effective heat transfer coefficients.
 
In a next step, we construct the device to be symmetric from a thermal point of view, because this symmetry halves the amount of free parameters necessary to describe thermal and charge transport. 
As there is no differentiation between top (encapsulation) and bottom (substrate) anymore, we can use a common effective heat transfer coefficient $h$ rather than distinguishing between $h_L$ and $h_R$.
To guarantee the desired symmetric distribution of the heat density, two conditions must be fulfilled.
Firstly, the layers left and right of the device center must have equal thermal conductivities $\kappa$.
With having reduced our thermal parameters to $\kappa$ and $h$, we can more clearly track their impact on the device temperature and performance.
Secondly, also the electric current density responsible for heat generation must be equal in the layers left and right of the device center.
This condition of balanced current halves the amount of necessary electrical parameters.
If realistic parameters are imposed for one layer, only parameters of the second layer require an adjustment to balance the current. 
In practice, the adjusted parameters are either the electron or the hole mobility in their respective transport layer or the offset between either the electron or the hole transport levels. 
A prototypical symmetric heating profile is shown in Figure~\ref{fig:setup}-b. We propose that such symmetric devices are ideal starting points to monitor the impact of subtle changes in the coupling between charge and heat transport.

\subsection{Material and geometry parameters.}
If not otherwise specified, we use the following parameters to setup our symmetric model OLED.
Each organic layer in our symmetric model device is 50~nm thick. 
The parameters for the electron transporting layer are inspired by Alq$_3$ and for the hole transporting layer by $\alpha$-NPB.
The offset between the electron transport levels is 0.3~eV, the offset between the hole transport level is the same. 
In each layer, the charge that is preferentially transported is considered with a charge carrier mobility $\mu_{high}$, whereas the charge of opposite polarity is considered with $\mu_{low}$. 
The mobilities are $\mu_{high}~=~6~\cdot~10^{-9}$ and $\mu_{low}~=~6~\cdot~10^{-10}$~\mobility, regardless whether electrons or holes are concerned. 
In the simulation, these mobility values correspond to the limit that the mobility attains for low electric fields and low charge carrier densities at room temperature. 
The relative dielectric permittivity $\epsilon_r$~=~3.5.
We disregard a direct thermoelectric coupling in our simulations by setting the Seebeck coefficient $S$ to zero.
Considering the typical values of $S~=~500~\mathrm{\mu V~K^{-1}}$ in organic semiconductors, the related thermal voltages $S \mathrm{\nabla} T$ are small compared to the applied voltage.
With this premise, the Seebeck coefficient has a negligible effect on the temperature and charge distribution and its disregard speeds up the simulations. 
We are using for both organic layers a mass density $\rho~=~2490~\mathrm{kg~m^{-3}}$ and a heat capacity $c_p~=~800~\mathrm{J~K^{-1}}$.
Note that mass density and heat capacity will not affect the steady state according to Eq.~(\ref{eq:contT}); rather, they affect the time scale in which steady state is reached.
Thermal conductivities of the organic layers realistically vary\cite{Reisdorffer2014} between $0.1~<~\kappa~<~1$~\thermalconductivity. They are inferior to the thermal conductivity  $0.8~<~\kappa~<~1$~\thermalconductivity~of ordinary glass,  ($\kappa$~=~1.022~\thermalconductivity~for a glass cover).


\section{Impact of thermal properties on the OLED operation}

\begin{figure}
	\centering
    \includegraphics{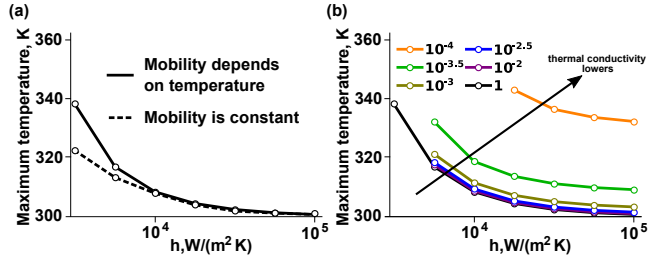} 
    \caption{(a) Difference between taking into account mobility dependence on temperature and having constant mobility. (b) Dependence of maximum temperature in the device on the thermal transfer coefficient $h$ for different heat conductivities (given in \thermalconductivity) for the device shown in Figure \ref{fig:setup} operated at 5~V. 
    For low values of $h$, particularly in combination with $\kappa$ values exceeding $10^{-3.5}$~\thermalconductivity~, such maximum temperatures cannot be given, because the generated heat cannot be fully dissipated from the device into its environment. Hence, the device heating never stops and the simulations fail to converge.}
\label{fig:I-Vandh-k-maxT}
\end{figure}

\subsection{Contributions to heating}
As a starting point, we explore how the generated heat is distributed across our symmetric model OLED operated at $V$~=~5V and at room temperature $T_{\mathrm{ambient}}$~=~300~K. 
Characteristic heat density distributions for three different values of $\kappa$ = $10^{-2}$, $10^{-3}$, and $10^{-4}$~\thermalconductivity 
are shown in Fig. \ref{fig:setup}-b.
Regardless of the values of $\kappa$ and $h$, these distributions are inherently symmetric with respect to the device center and consist of two distinct contributions. 
(i) The heat density features a sharp peak at the center. 
This contribution arises from the non-radiative electron-hole recombination that is localized at the organic-organic heterojunction. 
(ii) Joule heating is caused by the flow of charge carriers and stretches essentially throughout the entire cross-section. 
The largest amount of Joule heat is produced in the center, where the electric field that drives the current is highest. 
Correspondingly, the heat density drops from the center towards the contacts. 

The temperature distribution across the OLED is shown in Fig.~\ref{fig:setup}-c. Inherent to the symmetric profile of the generated heat, the temperature attains its maximum in the device center at the heterojunction. This maximum temperature depends on the value $\kappa$. However, only for a $\kappa$ as low as 10$^{-4}$~\thermalconductivity~
 a maximum temperature is established that significantly exceeds the ambient temperature (by 40~K).   

\subsection{Heat transfer coefficient vs. thermal conductivity}
To systematically clarify, how the temperature distribution $T(z)$ changes with $h$ and $\kappa$ for a given voltage, we performed simulations in which the values of the thermal conductivities and heat transport coefficients varied by several orders of magnitude. 
To learn which combinations of $\kappa$ and $h$ lead to realistic and significant temperature elevations (from 1 to 100~K above room temperature), we intentionally also incorporated values for $\kappa$ and $h$ that materials cannot necessarily attain.
The maximum temperature taken from the simulated temperature distributions is shown as a function of $h$ for several values of $\kappa$ in Figure \ref{fig:I-Vandh-k-maxT}.

Elevated maximum temperatures occur if either $h$ or $\kappa$ are small. Small values of $h$ and $\kappa$ correspond to a situation, in which we reduce the ability to dissipate heat from the OLED and, hence, store more heat inside the device.
The stored heat results in a strongly elevated temperature, as seen for $h<10^4$~\hcoefficient~ in Figures \ref{fig:I-Vandh-k-maxT}-a,b. Note that the temperature elevation is particularly driven by the thermally activated charge transport. 
As shown in Figure \ref{fig:I-Vandh-k-maxT}-a, the attained maximum temperature is considerably larger compared to a hypothetical situation with a temperature-independent mobility.
The thermal conductivities of the organic layers in the OLEDs, expected to be between 0.1 and 1 \thermalconductivity, have a negligible impact on the temperature; the associated temperatures in Figure~\ref{fig:I-Vandh-k-maxT}-b are indistinguishable.
Even when taking a seriously underestimated thermal conductivity of 0.01 W/mK, the corresponding temperature distribution in Fig.~\ref{fig:setup}-b (red line) exceeds the ambient temperature by less than 4~K. 
This rather unexpected, negligible impact of $\kappa$ roots in the fact that the typical thickness of the electrically active layers is not large enough to sustain a temperature gradient inside the layer. To see this, we develop the following rationale: 

For our device consisting of two electrically active organic layers, we determine the highest, $T_{\mathrm{max}}$, and the lowest temperature, $T_{\mathrm{min}}$, inside the OLED with the help of the steady state heat flow equation (Eq.~(\ref{eq:1Dheatequation})) for given operating conditions. These solutions $T_{\mathrm{max}}$ and $T_{\mathrm{min}}$ are then interpreted in terms of $\kappa$ and $h$.
The maximum temperature $T_{\mathrm{max}}$ serves us as an indicator, how well the materials and operating conditions support the heat dissipation from the device. 
Combining $T_{\mathrm{max}}$ with $T_{\mathrm{min}}$, we also get largest temperature difference with the device.
To ease a subsequent interpretation, we approximate the heat density distribution $H(z)$ with a centered uniform profile $H_{\mathrm{uni}} = const$. 
Choosing the extension of the uniformly heated region to be $2x$ (cf. Fig.~\ref{fig:setup}-b), the totally heat generated with this profile is $2 x H_{\mathrm{uni}}$.
The lowest temperature is adopted on the device surfaces, $T_{\mathrm{min}} = T(0) = T(L)$. 
Given the arc-like shape of the temperature distribution (cf. Fig.~\ref{fig:setup}-c), that is symmetric for our model devices, $T_{\mathrm{max}}$ is located in the center of the device $T_{\mathrm{max}}$~=~$T(L/2)$.
Hence, we get $T_{\mathrm{max}}$ and $T_{\mathrm{min}}$  from the general solution $T(z)$ of the steady state heat flow equation Eq.~(\ref{eq:1Dheatequation}) (cf. Supporting Information for full expression) at $z~=~0$ and $z~=~L/2$, respectively:
\begin{subequations}
      \begin{align}
      	T_{min}=&\, x H_{\mathrm{uni}}\, \frac{1}{h},                                             \label{eq:analtmin}\\
        T_{max}=&\, x H_{\mathrm{uni}}\, \left(\frac{1}{h}+\frac{L - x}{2\kappa}\right) .		 \label{eq:analtmax}
      \end{align}
      \label{eq:analt} 
\end{subequations}
Both temperatures are proportional to the total heat generated in the device, $2 x H_{uni}$.
The surface temperature $T_{min}$ is inversely proportional to $h$, but does not contain any dependence on the thermal conductivities of the electrically active layers.
The temperature $T_{max}$ is determined by two terms, one being inversely proportional to $h$ and one being inversely proportional to the thermal conductivity $\kappa$. 
This second term conveys two important insights:
Firstly, the thermal conductivity influences $T_{max}$ the lesser, the less concentrated the generated heat profile is, i.e., the smaller the difference $L-x$ between the thickness $L$ of the layers and the halfed extension $x$ of the heat density profile is.
Secondly, even in a best-case estimation, in which we insert realistic values $L$~=~150~nm and $\kappa$~=~0.1~\thermalconductivity, $(L-x)/\kappa$ remains with $\leq$ 1.5~\thermalconductivity~at least an order of magnitude smaller than the 1/$h$ contribution.
The thermal conductivity of the organic materials would non-negligibly contribute to $T_{max}$ if the organic films were at least a factor of ten thicker.

At this point, the irrelevance of the thermal conductivity $\kappa$ for thermal transport in realistically thin organic films has two important consequences.
Firstly, the actual bottleneck for heat dissipation is the combined thermal transfer between the contacts and the exterior, i.e., the thermal conductivities of layers being not electrically active and the associated heat transfer coefficients.
Vice versa, the layers that are responsible for electric transport are not relevantly involved in heat dissipation. 
This implies that the thermal and electrical properties can be optimized \textit{independent} from each other in complementary regions of the device to reach a desired performance. 
Secondly, the temperature within the organic layers is uniform, i.e., $T(z)$ = $T$ = const.
Then, $T$ can be immediately related to the total amount of generated heat per unit area, $H_{tot}$, via the heat balance equation $h_L T + h_R T = H_{tot}$:
\begin{equation}
	T=\frac{H_{tot}}{h_L+h_R}.
    \label{eq:onlyh}
\end{equation}
Note we do not require a symmetric device to establish this relation. 
$H_{tot}$ equals $H_{uni}~x$ in the case of a rectangular heating profile.
The heat transfer from the contacts to their associated device surfaces is fully accounted for in the effective heat transfer coefficients $h_L$ and $h_R$.

\subsection{Impact of heat transfer $h$ on the current-voltage characteristics}

With having established that the thermal conductivity has a negligible impact on the device temperature, we turn to a closer inspection of the role of $h$.

If the maximum temperature is already markedly elevated, as seen in Figure \ref{fig:I-Vandh-k-maxT} for the smaller $h$, small changes in $h$ cause a large change in the temperature.
This finding is potentially relevant if one considers to operate the OLED considerably above room temperature, as such elevated temperatures promote the electric conduction and radiative recombination. Subtle changes in the OLED layout, e.g., in terms of the thickness of the encapsulation layer, may strongly alter the temperature in the device, possibly even pushing the device in the regime of insufficient cooling and, concomitantly, into self-heating and thermal run-away.

In a next step we explore how $h$ affects the electric current and heating for different applied voltages.
To obtain a current-voltage characteristic, we computed the current for each external voltage anew starting out from an operating temperature equal to the ambient temperature. Hence, the computed I-V curves correspond to electric measurements that allow the device to sufficiently cool between consecutive current acquisitions.

\begin{figure}
\centering
\includegraphics{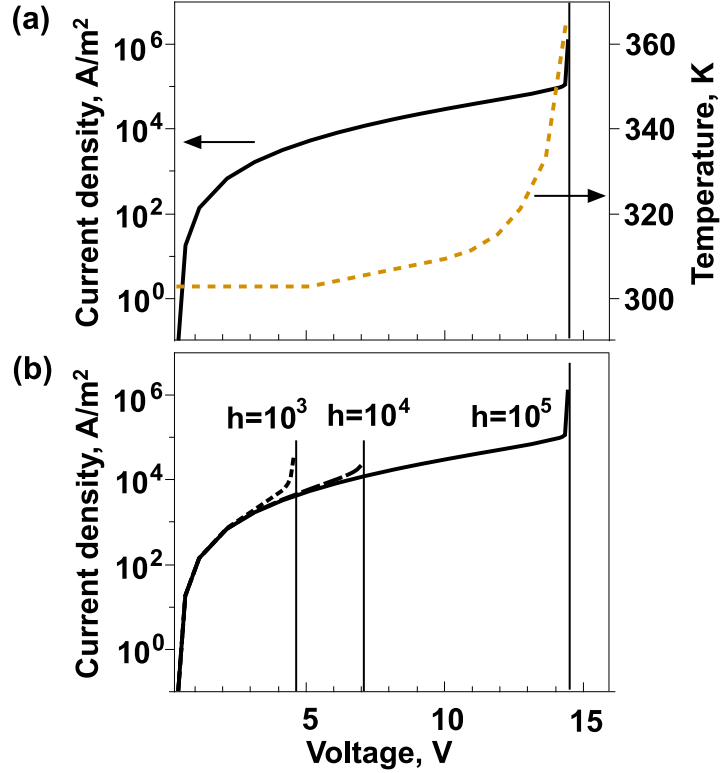} 
\caption{(a)~Current (solid line) and temperature (dashed line) against voltage for the device shown in Figure \ref{fig:setup} for $\kappa$~=~\thermalconductivity~and $h$~=~10$^{5}$~\hcoefficient~operated at room temperature $T$~=~300~K. The voltage at which the current diverges is indicated with a vertical line. Beyond this voltage, the temperature inside the device rises too quickly to allow a full heat dissipation and prevents a stable device operation (vicious cycle). (b)~Current-voltage characteristics for heat transfer coefficients $h$ between 10$^{3}$ and 10$^{5}$~\hcoefficient. The characteristics coinside at small voltages. The better the device is cooled (the larger $h$), the larger is the voltage, beyond which the current diverges (indicated by vertical lines).}
\label{fig:I-V}
\end{figure}

Figure~\ref{fig:I-V}-a shows the current of the symmetric model OLED with $\kappa$~=~1~\thermalconductivity~and $h$~=~10$^{5}$~\hcoefficient~operated at room temperature $T$~=~300~K a function of the applied voltage.
As the voltage is increased, the current and the temperature increase. 
Remarkably, also the thermal runaway process can be readily identified in this plot as an abrupt end of the recorded current-voltage curve that is preceeded by a sharp kink. The corresponding voltage is indicated by a vertical line in Figure~\ref{fig:I-V}-a.
Beyond this operating voltage, it is not possible to establish a stable current, because the heat is not sufficiently removed from the device anymore. The better the device is cooled, i.e., the larger the value of $h$, the larger is the critical voltage up to which the OLED operates. 

This impact of cooling is illustrated in Figure~\ref{fig:I-V}-b, that shows the evolution of the current density - voltage relation for the same charge transport parameters upon increasing the heat transfer coefficient $h$ from $10^3$ to $10^5$ \hcoefficient. 
The former value corresponds to the least efficient and the latter value to the most efficient cooling.
Again, the operating voltage in which we find the maximum current density and beyond which the current fails to converge is marked with a vertical line. 
All three current voltage characteristics coincide as long as the OLED is operated safely below 4.5~V.
For the least efficient cooling considered, $h$~=~10$^3$~\hcoefficient, the current at 4~V is an order of magnitude larger than the current for a much more efficient cooling ($h \leq 10^4$ \thermalconductivity). Already small fluctuations in the operating voltage may cause a thermal runaway.
The current voltage characteristics do not only reflect the material and geometrical properties of the active layers, but also account for the properties of the exterior layers and, in effect, operating conditions, e.g., whether convective cooling applies.

\subsection{Extraction of heat transfer coefficients from experiment}

We have developed a rationale (Eq.~(\ref{eq:onlyh})), according to which the temperature $T$ in the electrically active layers is determined by the ratio between the total heat generated per unit area, $H_{tot}$, and the effective heat transfer coefficients $h_{(L/R)}$ between the electrically active region and the exterior with ambient temperature. 
With this relation at hand, the question arises whether it is possible to obtain the totally generated heat density $H_{tot}$ directly from experiment to get an estimate for either $h_L$ and $h_R$ for a known device temperature $T$ or $T$ for known $h_L$ and $h_R$.
Such an estimation is valid as long as the heat transported away from the electrically active region does not influence the temperature inside the this region and preserves a uniform temperature distribution. 

The total heat density $H_{tot}$ consists of the Joule heat $H_J$ and the heat $H_{\mathrm{rec}}$ due to non-radiative recombination.
Knowing the current voltage characteristics, $I(V)$, the Joule heat per unit area $S$, $H_{J}$, is readily given by
\begin{equation}
		H_{J}= \frac{I V}{S}. \label{eq:JouleHeating} 
\end{equation}
The recombination heat per unit area $S$, $H_{\mathrm{rec}}$, equals
\begin{equation}
		H_{\mathrm{rec}} = \frac{(1-\eta_\mathrm{i}) I E_R(eV)}{S}. 		\label{eq:RecombinationHeating}
\end{equation}
Therein, the coefficient $1-\eta_i$ ensures that exclusively triplet exciton states are considered for the heating and $E_R(eV)$ is the recombination energy, measured in electron-volts. 

Inserting Eqs.~(\ref{eq:JouleHeating},\ref{eq:RecombinationHeating}) into the analytic equation Eq.~(\ref{eq:analtmax}) for the maximum temperature, we arrive at:
\begin{equation}
	T_{\mathrm{max}}=\frac{ V+(1-\eta_i) E_R\left(eV\right) }{h_L+h_R} \frac{I}{S}.
    \label{eq:TmaxAsAFunctionOfIVErec}
\end{equation}

We subjected this relation for $T_{\mathrm{max}}$ to a consistency check to reveal whether we can safely exclude a possible feedback from the heat transported through the encapsulation and substrate layers on the temperature inside the electrically active layers.
To this aim, we compared the value $T_{\mathrm{max}}$ extracted from the full simulation with the value obtained from Eq.~(\ref{eq:TmaxAsAFunctionOfIVErec}), in which we inserted only the current densities obtained from  the simulations. For a realistic thermal conductivity of 0.5~\thermalconductivity~with an voltage of 5 V applied, we calculated $T_{\mathrm{max}}$ from Eq.~(\ref{eq:TmaxAsAFunctionOfIVErec}). 
Eq.~(\ref{eq:TmaxAsAFunctionOfIVErec}) reproduces the exact temperature values for a large range of heat transfer coefficients $h$; for $h_L + h_R \geq 10^{4}$~\hcoefficient~the relative error in predicted temperatures remains below 1~\%. 
Only when $h$ approaches values for which the device starts to overheat, either a deviation ($4\cdot 10^3 \leq h_L + h_R \leq 10^4$ \hcoefficient) or thermal runaways are obtained ($h_L + h_R \geq 4\cdot 10^3$ \hcoefficient). 
Details are shown in the Supporting Information.

\section{Conclusions}

We investigated how the thermal conductivities of the electrically active organic layers and the heat transfer between the electrically active region and the ambient environment determine the temperature distribution inside OLEDs under operating conditions.
To this aim, we established a macroscopic, three-dimensional drift-diffusion based simulation that incorporates charge transport, thermal transport and their mutual coupling for a given OLED stack.
In particular, we account for the feedback mechanism of self-heating, that occurs when any dissipation of heat within electrically active layers boosts their electric conductivity which, in turn, enhances the generation of heat.
This model allows us to monitor the time-dependent and steady-state behavior of the temperature, the charge carrier densities, the current densities, and local electric fields.

The OLED reaches steady state operation only if the heat is sufficiently well removed from the device to counter-balance self-heating. 
If either the thermal conductivities or the heat transfer coefficients fall below a certain limit, the blocked outflow of heat promotes a steep increase in the temperature inside the OLED with rising applied voltage. 
For such a choice of the thermal parameters, any other change in the OLED setup or any fluctuations in operating conditions can trigger the thermal runaway process.

The thermal conductivity of realistically thin organic layers only marginally affect the temperature in the device; in fact, the temperature adopted in an electrically active organic layer is essentially constant. 
The temperature profile across the entire stack is determined by the thermal properties of the encapsulation layers, the substrate layers, and the heat transfer coefficient associated to heat exchange between the outermost OLED surfaces and the ambient environment.
Self-heating arising from an inadequate heat dissipation from the heat source towards the exterior, e.g., due to a limited choice in terms of the substrate, can neither be counter-balanced by a larger thermal conductivity, nor by thicker organic films. 

The layers responsible for the major electrical effects differ from the layers in which the relevant thermal effects occur, despite the fact that charge and heat transport strongly couple in the electrically active layers. 
From a practical point of view that means that the charge transport can be considered and optimized assuming a fixed temperature (e.g., in Kinetic Monte Carlo simulations). The temperature distribution across the entire stack can be, in turn, provided by solving the heat transport equation for the layers outside the electrically active substack.

\section{Acknowledgement}
This work was funded by the Austrian Climate and Energy Fund (KLIEN) and the Austrian Research Promotion Agency (FFG) through the project “ThermOLED” [FFG No. 848905]. The authors would also like to acknowledge the use of HPC resources provided by the ZID of the Graz University of Technology.

\clearpage
\bibliographystyle{apsrmp4-1} 
\bibliography{New_citation.bib}

\end{document}


\textsf{\textbf{Impact of thermal transport parameters on the operating temperature of OLEDs}}\\[0.3cm]
\textsf{G. Krikun and K. Zojer}\\[0.3cm]
\textit{Institute of Solid State Physics and NAWI Graz, Graz University of Technology, Graz, 8010, Austria}

\addtocounter{figure}{5}
\addtocounter{equation}{11}

\section*{Supporting information.}

\subsection*{General analytic solution for the steady-state 1D heat equation}

Utilizing a thermal conductivity function, in which a thermal conductivity $\kappa_j$ is assigned to each layer $j$ located along the $z$-axis (cf. Fig.~3), we can formulate a heat equation for the whole device.
The general solution of the differential equation (5) is given by a Greens function $G(z,z_0)$ that solves Eq.~(\ref{eq:GreenFunctionEquation}). 

\begin{equation}
-\left( \kappa(z) G_z'(z,z_0)  \right)_z' = \delta(z-z_0)
\label{eq:GreenFunctionEquation}
\end{equation}

If one multiplies Eq.~(\ref{eq:GreenFunctionEquation}) by $H(z_0)$ and integrates it with respect to $z_0$, one obtains the following equation: 
\begin{equation}
-{\left(\kappa (z) \left(\int_0^L dz_0 G(z,z_0) H(z_0) \right)_z' \right)}_z' = H(z).
\label{eq:GreenFunctionHeatEquation}
\end{equation}

Therefore, the term in the inner bracket of Eq.~(\ref{eq:GreenFunctionHeatEquation}) should be equal to the $T(z)$ in the presence of the heat distributed as $H(z)$. 

If one returns to the equivalent notation with several layers and imposes the preservation of heat flux across the interfaces, one arrives at the equation: 
\begin{equation}
T(z) = \sum_{j'=1}^{N} \int_{z_{j'-1}}^{z_{j'}} H(z_0) G(z,z_0) dz_0.
\label{eq:GreenFunctionSolution}
\end{equation}

Where $z_{j'}=\sum_{i=1}^{j'} L_i$ and $L_{i}$ refers to the thickness of the individual layers.
The Greens function $G(z,z_0)$ can then be conveniently written using functions $W^L(z)$ and $W^R(z)$ for each layer $b$:

\begin{subequations}
    \begin{align}
    W_b^L(z) =& \frac{1}{h_L}+\sum_{i=1}^{b-1} \frac{L_i}{\chi_i} + \frac{-\left( \sum_{i=1}^{b-1} L_i \right) + z}{\chi_b} \label{eq:WL}\\
    W_b^R(z) =& \frac{1}{h_R}+\sum_{i=b+1}^{N} \frac{L_i}{\chi_i} + \frac{\left( \sum_{i=1}^{b} L_i \right)  - z}{\chi_b}. \label{eq:WR}\\
    \end{align}
    \label{eq:Wfunction}
\end{subequations}

Then, the Greens functions $G^{L}(z,z_0)$ for $z<z_0$ and $G^{R}(z,z_0)$ for $z>z_0$ read:

\begin{subequations}
    \begin{align}
    G_{j,j'}^L =& \frac{1}{\sum_{i=1}^{N} \frac{L_i}{\chi_i}} W^L_{j}(z) W^R_{j'}(z_0), \quad \mathrm{for~} z<z_0 \label{eq:GL}\\
    G_{j,j'}^R =& \frac{1}{\sum_{i=1}^{N} \frac{L_i}{\chi_i}} W^R_{j}(z) W^L_{j'}(z_0), \quad \mathrm{for~} z>z_0. \label{eq:GR}
    \end{align}
    \label{eq:greenfunction}
\end{subequations}

While $j$ and $z$ are the coordinates of the layer where one calculates temperature, $j'$ and $z_0$ correspond to that layer from which one want to calculate contribution to the temperature (cf.~Figure~3). 
To obtain temperature distribution in certain layer $j$ at point $z$ for given heat distribution $H(z_0)$ according to Eq.~(\ref{eq:GreenFunctionSolution}), one should integrate $H(z_0)$ weighted with the corresponding Greens function $G_{j,j'}(z,z_0)$ with respect to $z_0$ and sum the contributions from all layers $j'$ . 

\begin{figure}[h]
    \centering
    \includegraphics{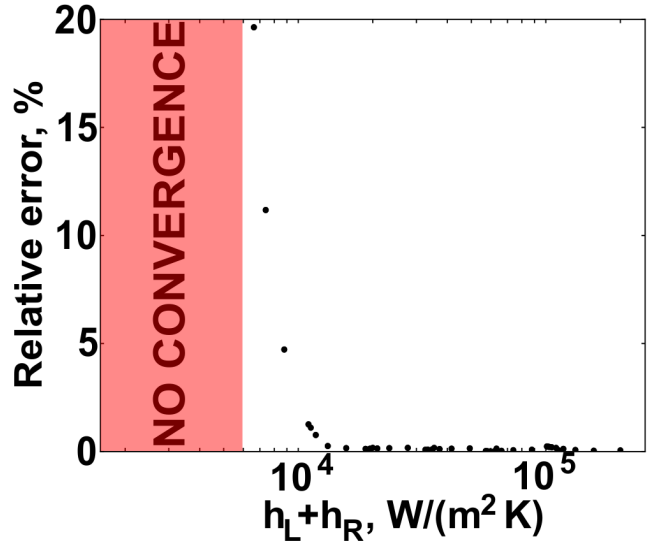}
    \caption{Validity of the maximum temperature predicted by Eq.(11) as a function of the total heat transfer coefficient $h_L + h_R$. Shown is the relative error $\frac{|T_{max,an} - T_{max,sim}| }{T_{max,sim}-300K} - 1$ formed between $T_{max,sim}$ deduced from the full simulation and $T_{max,an}$ predicted from Eq.(11) using only the current obtained from the simulation. In the red shaded region, located at small $h_L + h_R$, the device overheats so that the simulations do not converge.}
    \label{fig:connection_to_experiment}
\end{figure}
\subsection*{Testing the consistency of the analytical model with simulation results.}

The maximum temperature predicted by the simplified Eq.~(11) corresponds well to the prediction of the full simulation for a large range of heat transfer coefficients.
Figure~\ref{fig:connection_to_experiment} shows the relative error,
$\frac{|T_{max,an} - T_{max,sim}| }{T_{max,sim}-300K} - 1$, between $T_{max}$ either predicted on the basis of Eq.(11) and the current obtained from the simulation ($T_{max,an}$) and $T_{max}$ directly obtained from the simulations ($T_{max,sim}$).

For $h_L + h_R < 4\cdot10^3$~\hcoefficient~, $T_{max,sim}$ cannot be determined, because thermal runaway prevents the convergence to a steady state (red shading).
Solely closely above the thermal runaway region, $h_L + h_R < 10^4$~\hcoefficient~, the temperature predicted by Eq.(11) deviates from the simulation. The prediction of Eq.(11) fails for these heat transfer coefficients, because the temperature distribution in the device is not uniform.

\subsection*{Feedback between heat transport and device temperature.}

Here we intend to identify for which $h$ and $\kappa$ it is safe to assume that the heat transport away from the electrically active region does not affect the temperature of this region. 
For situations in which this assumption is invalid, we want to gather the symptoms of the electro-thermal feedback.
We check for the presence of an electro-thermal feedback by comparing the temperature distribution $T_{sim}(z)$ from the full simulation and $T_{an}(z)$ from solving the steady state heat transport equation, Eq.~(5).
Eq.~(5) is inherently unable to account for a possible feedback of the heat transport on the charge transport and the heat generation profile.
Hence, we can interpret any deviation of $T_{an}(z)$ from $T_{sim}(z)$ as a manifestation of a feedback.

Such a comparison is particularly feasible for our symmetric model OLED.
As the temperature distribution is symmetric, the largest temperature $T_{max}$ is always established in the device center (cf. Fig.~2-c). Rather than accounting for the entire profiles $T(z)$,
it is sufficient to compare the predicted maxima $T_{max,an}$ and $T_{max,sim}$.
For sufficiently large $\kappa$ and $h$, we expect that the heat flow is large enough to prevent any heat accumulation in the electrically active region. 
Hence, the heat density profile $H(z)$ is independent of $\kappa$ and $h$. Guided by Fig.~4-b, we encounter this situation for $\kappa = 1$~\thermalconductivity~ and $h = 10^5$~\hcoefficient~ at an operating voltage of 5~V.  
The simulated heat density profile $H(z)$ retrieved for these thermal parameters gives rise to the total heat $H_{tot} = \int_0^L dz H(z)$. 
Since this $H(z)$ is solely governed by the electric properties and operating conditions, it will now serve as the input for Eq.~(5) to inspect $T_{an}$ for other values of $\kappa$ and $h$.

\begin{figure}
    \centering
    \includegraphics[width=0.75\textwidth]{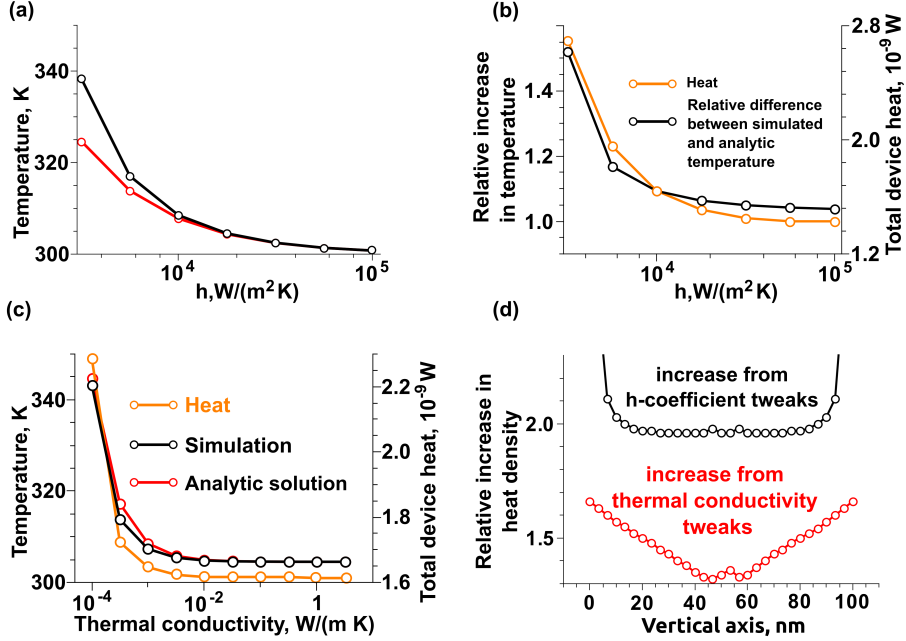}
    \caption{(a) Maximum temperature $T_{max,sim}$ (black) and $T_{max,an}$ (red) in the device as a function of the heat transfer coefficient $h$ at $\kappa$=1~\thermalconductivity. $T_{max,sim}$ are temperature values extracted from simulations of the model OLED operated at 5~V. $T_{max,an}$ are temperature values calculated from the analytical solution of Eq.~(5) using the heat generation profile obtained for $h=10^5$~\hcoefficient. (b) Relative increase in temperature, $\frac{T_{max,sim}}{T_{max,an}}-1$, (left axis) compared to the total device heat $H_{tot}$ (right axis) as a function of the h-coefficient. (c) Maximum temperature $T_{max,sim}$ (black) and $T_{max,an}$ (red) in the device (left axis) and $H_{tot}$ (right axis) as a function of thermal conductivity at h = $10^{4.25}$~\hcoefficient. (d) Local changes in heat density due a change in a thermal parameter. $\frac{H_{h}(z)}{H_{h_{ref}}(z)}-1$ monitors the relative change in heating density when going from $h = 10^{3.5}$ to $10^5$~\hcoefficient at $\kappa$ = 1~\thermalconductivity (black curve) and $\frac{H_{\kappa}(z)}{H_{\kappa_{ref}}(z)}-1$ the relative change in heat density in going from
        $\kappa = 1$ to $10^{-4}$~\thermalconductivity~for $h$ = $10^{4.25}$~\hcoefficient (red curve).}
    \label{fig:final4plots}
\end{figure}

Figure~\ref{fig:final4plots}-a shows the evolution of $T_{max,an}$ and $T_{max,sim}$ for varying heat transfer coefficients $h$ at $\kappa = 1$~\thermalconductivity. 
As long as $T_{max}$ remains smaller than 310~K, $T_{max,an}$ (red) and $T_{max,sim}$ (black) are indistinguishable.
Below $h \approx 10^4$~\hcoefficient, the simulations reveal a markedly larger temperature as one would expect from the analytical solution $T_{max,an}$. 
Not only the two temperatures $T_{max,sim}$ and $T_{max,an}$, but also the trend in the deviation between the two temperatures, cast for convenience into the expression $\frac{T_{max,sim}}{T_{max,an}}-1$ in Fig.~\ref{fig:final4plots}-b, is directly related to the total amount of heat $H_{tot}$ (orange) produced in the device.
Hence, the elevated temperature $T_{max,sim}$ is the symptom for the electro-thermal feedback between the transported heat and the heat density profile that causes self-heating.

Note that the equality between $T_{max,sim}$ and $T_{max,an}$ does not necessarily imply the absence of electro-thermal feedback. 
To illustrate that, Fig.~\ref{fig:final4plots}-c shows $T_{max,sim}$ and $T_{max,an}$ as a function of $\kappa$ at constant $h = 10^{4.25}$~\hcoefficient. 
At the first glance, $T_{max,sim}$ and $T_{max,an}$ coincide even for lowest probed $\kappa = 10^4$~\thermalconductivity.
However, a closer inspection of the situation at such low $\kappa$ reveals that there is nevertheless an electro-thermal feedback and, hence, heat accumulation inside the electrically active layers.
This accumulation of heat predominantly changes the \textit{shape} of the heat density profile $H(z)$.
To visualize how the heat density $H(z)$ differs in each position $z$ in the electrically active region between $10^{-4}$~\thermalconductivity and $\kappa$ = 1~\thermalconductivity, we plot the corresponding relative change of heat density, $\frac{H_{\kappa}(z)}{H_{\kappa_{ref}}(z)}-1$ (red curve) in Fig.~\ref{fig:final4plots}-d.
This change in heat density does not occur uniformly across the active layers. 
Rather, regions are heated the stronger the closer they are to the surface.
Due to the arc-shaped temperature distribution across the within the active layers, a lowering of $\kappa$ causes the temperature to rise mainly in the center of the device and, to a much lesser extent, near the outer surfaces. 
While the temperate elevation increases the recombination and the electric conductivity at the device center, the electrical conductivity near the surfaces considerably lesser raised.
Since Joule heating is more pronounced in the regions with lower electrical conductivity, more heat is produced near the surface compared in the center.

For comparison, we also show that shape of the heat density profile $H(z)$ is essentially preserved when comparing excellent ($h$ = $10^{5}$~\hcoefficient) and inferior ($10^{3.5}$~\hcoefficient) heat transfer with the environment.
Fig.~\ref{fig:final4plots}-d (black curve) shows the corresponding relative change of heat density, $\frac{H_{h}(z)}{H_{h_{ref}}(z)}-1$.
Throughout the active layers, this change is practically uniform except for a sudden increase in the close vicinity of the outer surfaces. The overall temperature profile is governed by these extended regions of uniform heating; the very narrow surface regions do not have a perceivable impact.